%% file: conference_041818.tex
\documentclass[conference]{IEEEtran}
\IEEEoverridecommandlockouts
\usepackage{cite}
\usepackage{amsmath,amssymb,amsfonts}
\usepackage{algorithmic}
\usepackage{graphicx}
\usepackage{textcomp}
\usepackage{xcolor}
\usepackage[keeplastbox]{flushend}
\usepackage{booktabs}
\usepackage{lipsum}
\input {latex-base-files/packages}
\usepackage{subfig}
\usetikzlibrary{decorations.pathreplacing}
\usepackage{caption}

\newcommand{\mtx}{\boldsymbol}

\DeclareMathOperator*{\argmax}{arg\,max}

\def\BibTeX{{\rm B\kern-.05em{\sc i\kern-.025em b}\kern-.08em
    T\kern-.1667em\lower.7ex\hbox{E}\kern-.125emX}}
\begin{document}

\title{Performance Evaluation of OTFS Over Measured V2V Channels at 60\,GHz}

\author{\IEEEauthorblockN{Thomas Blazek}
  \IEEEauthorblockA{Institute of Telecommunications\\
    TU Wien,
    Vienna, Austria \\
    thomas.blazek@tuwien.ac.at}
  \and
  \IEEEauthorblockN{Danilo Radovic}
  \IEEEauthorblockA{Institute of Telecommunications\\
    TU Wien,
    Vienna, Austria \\
    danilo.radovic@tuwien.ac.at}
}

\maketitle

\begin{abstract}
  This paper presents an analysis of the Orthogonal Time Frequency Space (OTFS)
  modulation scheme when applied to realistic vehicular channel situations. OTFS
  modulates symbols in delay-Doppler domain, hoping to exploit diversity in
  both. The penalty for doing this is the requirement of complex interference
  cancellation equalizers, as this domain incurs a strong amount of intercarrier
  and intersymbol interference. We conduct this analysis using measured millimeter wave vehicular channels, and we assume
  typical physical layer settings for a performance analysis. Our results show that there is a challenging trade-off between channel conditions that are easy to equalize and channel conditions that allow OFTS to exploit the two-dimensional diversity. In the first case we observe a good overall performance that is barely enhanced by employing OTFS. In the second case performance gain through OTFS is visible, yet with a
  bad overall performance.
\end{abstract}


\section{Introduction}
Vehicular communications pose unique challenges to wireless communications. Due
to the openness of space and high mobility of the observed channel, large delay- and
Doppler spreads are to be expected \cite{Blaz1809:Model,Mecklenbrauker2011}. The
current generation of \ac{V2X} communication protocols struggle with these
situations due to difficult channel estimation. For
future generations, different approaches have been proposed to mitigate this.
IEEE 802.11bd inserts a mid-amble to enable easier channel
estimation, while 5G proposes different modes to tackle with different
scenarios~\cite{naik2019ieee}.

However, many of the current approaches still base their physical layer solutions on \ac{OFDM}, which
has its own limitations. Specifically, the format is based on a per-subcarrier
one-tap equalization. While this provides  simplicity in equalization of static channels, channel conditions
in frequency domain are prone to fast changes, which is problematic for massive
\ac{MIMO} and high mobility applications. Recently, the authors of \cite{monk2016otfs} have
proposed a two-dimensional modulation scheme, \ac{OTFS}. The idea has been
expanded in \cite{hadani2017orthogonal}. In this scheme, the symbols are spread
out in delay- and Doppler domain, to enable exploitation of diversity in both.
On the one hand, the presented approach does raise the required complexity by necessitating complex iterative decoders
\cite{viterbo-decoder}. On the other hand, two dimensional modulations schemes
have been shown to improve throughput \cite{zemen2018iterative}, and result in
relatively sparser channels \cite{TUW-279699} due to the high dimensionality of
the channel. Furthermore, \ac{OTFS} has been
projected to better deal with highly time-variant channels~\cite{murali2018otfs} and
scale well to massive \ac{MIMO}~\cite{ramachandran2018mimo}. Part of the
projected gains are based on the promise that the observed channel will be sparse in the \ac{OTFS} domain.
\subsection{Our Contribution}
As promising as this schemes is, careful analysis of the real-world
applicability and potential has to be conducted. While OTFS is currently a strong research topic (see e.g. \cite{raviteja2019embedded,8746382,khammammetti2018otfs}), no analysis with
real-world channels in the loop has currently been published. In this work, we 
evaluate the performance of \ac{OTFS} when presented with actual \ac{mmWave}
measurements that represent a vehicular urban overtaking scenario. To achieve
this, we take channel measurements conducted in \cite{zochmannmeasured}, and
combine them with an \ac{OTFS} simulator setup.
The given measurements were conducted at $60\,$GHz, and measured a urban
scenario. The results were shown to exhibit sparse channel properties
\cite{blazek2018millimeter}. Thus, they are an ideal candidate to evaluate
\ac{OTFS} performance. 
We present a performance analysis of \ac{OTFS} using an iterative \ac{MP}
decoder presented in \cite{viterbo-decoder}. The performance analysis is
conducted against channel sounding measurements that were done at $60\,$GHz for
an urban overtaking scenario. We consider typical communication system
transmission parameters. Our results show that it is not trivial to establish
gains from an \ac{OTFS} system. Parameter settings that lead to sparse delay
domains are not spread out over the Doppler-domain. Conversely, settings that do
diversify over Doppler see very dense delay channels. Hence, there is a
trade-off between an overall sparse channel, and spreading the channel to gain
from diversity.
The next section presents the OTFS system model. In section III, we present the
channel measurements, as well as our methodolgy to adapt them for the
simulations. Section IV presents our simulation results.
\section{System Model}

\subsection{OTFS Modulation}
\begin{figure}
  \centering
  \begin{tikzpicture}
    \begin{scope}[gray,semithick]
      \foreach \x in {1,...,6}2
      \draw (\x,1) -- (\x,4);
    \end{scope}
    \begin{scope}[semitransparent,semithick]
      \foreach \y in {1,1.5,...,4}
      \draw (1,\y) -- (6, \y);
    \end{scope}
    
    \draw [decorate,decoration={brace,amplitude=10pt},xshift=-4pt,yshift=0pt]
    (1,1) -- (1,1.5) node [black,midway,xshift=-0.6cm] 
    {\footnotesize $\Delta  \tau$};

    \draw [decorate,decoration={brace,mirror,amplitude=10pt},yshift=-3pt]
    (1,1) -- (2,1) node [black,midway,yshift=-0.5cm] 
    {\footnotesize $\Delta  \nu$};
    \draw [decorate,decoration={brace,amplitude=10pt},yshift=3pt]
    (1,4) -- (6,4) node [black,midway,yshift=0.5cm] 
    {\footnotesize $M\Delta  \nu$};
    \draw [decorate,decoration={brace,amplitude=10pt},xshift=3pt]
    (6,4) -- (6,1) node [black,midway,xshift=0.8cm] 
    {\footnotesize $N\Delta  \tau$};
    \draw (1.5,3.75) node {$a_{1,1}$};
    \draw (2.5,3.75) node {$a_{1,2}$};
    \draw (1.5,3.25) node {$a_{1,2}$};
    \draw (2.5,3.25) node {$a_{2,2}$};
    \draw (1.5,2.75) node {$\vdots$};  
    \draw (3.5,3.75) node {$\cdots$};  
    \draw (3.5,2.75) node {$\ddots$}; 
  \end{tikzpicture}\caption{Example of an OTFS grid. }\label{fig:lattice}
\end{figure}
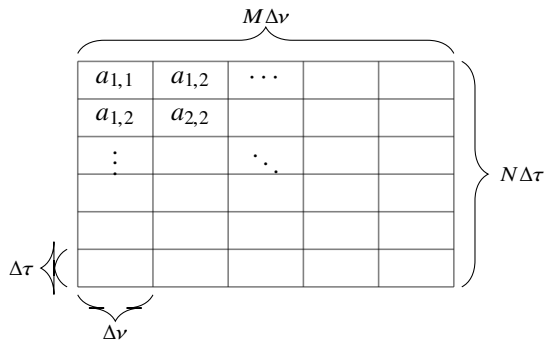
\ac{OTFS} defines a delay-Doppler grid given by the lattice
 that consists of the cartesian product of the two dimensions \cite{hadani2017orthogonal}
\begin{equation}
  \label{eq:delay-doppler-grid}
  \Lambda^\bot = \left\{\big(n\Delta\tau, m\Delta \nu\big) : n \in [1,N]\,, m\in[1,M]\right\}.
\end{equation}
Fig. \ref{fig:lattice} illustrates this grid. $\Delta\tau$ defines the sampling period
of the system, while $\Delta\nu$ is the lowest resolvable Doppler shift. $N$ and
$M$ are the total number of symbols in delay and Doppler-domain respectively.
\ac{OTFS} requires that delay- and Doppler grid resultions have to be related via 
\begin{equation}
  \label{eq:condition}
  \Delta\nu = \frac{1}{NM\Delta\tau}.
\end{equation}
Each element of this grid is assigned a symbol $a_{n,m}$, e.g. from a QAM
alphabet $\mathbb A$.

We now
denote the matrix of transmit symbols $\mtx X \in \mathbb R^{N\times M}$. There
are multiple ways to transmit this block. One way is to transform  the Doppler
domain to time-domain, resulting in a data block in fast (delay) and slow (time)
domain. This block can be transmitted in an appropriately
interleaved fashion. Alternatively, the matrix is transformed into time-frequency domain using the \emph{inverse symplectic
  discrete fourier transform} (IDSFT)
\begin{equation}
  \mtx U[f,t] =\frac{1}{\sqrt{NM}}\sum_{n=0}^{N-1}\sum_{m=0}^{M-1}\mtx X[n,m]e^{-j2\pi(tm/M - fn/N)}.
\end{equation}
$f$ and $t$ denote indices in frequency and (slow) time domain, and relate to
physical times $t'$ and $f'$ via
\begin{align}
  \label{eq:transforms}
  t' =& \frac{ t}{M\Delta\nu} + T_0, \\
  f' =& \frac{f}{N\Delta\tau} + F_0.
\end{align}
$F_0$ and $T_0$ refer to the carrier frequency and start time of transmission.
The symbol block $\mtx U$ can for example be transmitted using an \ac{OFDM} frontend. We now assume that the
impulse responses within single subcarriers are reasonably flat. This can always
be achieved by using a \ac{CP} in conjunction with the \ac{OFDM} transmission.
Then, given a block of channel transfer functions $\mtx H[f,t]$, the received block
$\mtx  R[f,t]$ equals \cite{zemen2018iterative}
\begin{equation}
  \label{eq:rx}
  \mtx R = \mtx H \odot \mtx U, 
\end{equation}
where $\odot$ denotes the (element-wise) Hadamard product. The received
block can be transformed back to a delay-Doppler representation $\mtx Y[n,m]$.

Alternatively, the matrix  $\mtx H$  can be represented in delay-Doppler domain
as $\mtx S_h$ \cite{TUW-279699}, and the input-output relation can be described
via the twisted convolution \cite{zemen2018iterative}
\begin{equation}
  \label{eq:twisted}
  \mtx Y = \mtx S_h \star \mtx X.
\end{equation}
The received block can then be equalized. Due to the large amount of \ac{ISI}
that this scheme incurs, we resort to a iterative decoding scheme.
We use the \acf{MP} algorithm presented in \cite{viterbo-decoder}.
The goal is to obtain the posterior estimate
\begin{equation}
  \label{eq:hat}
  \hat{\mtx X} = \argmax_{\mtx X \in \mathbb A^{N\times M}} \Pr(\mtx X | \mtx Y, \mtx H).
\end{equation}
This maximization is applied over the whole symbol matrix. In
\cite{viterbo-decoder}, this maximization is approximated by a
element-by-element optimization. This element-wise optimization is updated and
iterated over the whole matrix, until convergence or an iteration limit is
reached.
\subsection{Simulation Setup}
For the performance evaluation in Tab. \ref{sec:results}, we assume our simulation
setup as follows. For each entry of $\mtx X$, we draw a random symbol from a QAM
alphabet. We then transmit the channel as
$\mtx  Y =  \mtx S_h \star \mtx X + \mtx N$,
where every entry of $\mtx N$ is a zero mean complex Gaussian. The variance will
be set to enforce a given \ac{SNR}, defined as the average bit energy over the noise power $E_b/N_0$.
In this paper, we assume
perfect \ac{CSI}, i.e. we assume to know $\mtx H$.
\subsection{System Parameters}
\begin{table}[t]
	\caption{Communication System Paramters}\label{tab:system-parameters}
	
	\centering
	\begin{tabular}{r|l}\toprule
		Parameter		&  Value \\ 
		\midrule
		Center frequency 	& $60$\,GHz \\
    System bandwidth & $5$, $40$, $120$\, MHz \\
		Number of subcarriers	& $64$ \\ 
		Number of Doppler samples & $2$, $8$, $64$\\
		\bottomrule
	\end{tabular} 
\end{table}
The system performance strongly depends on the choice of transmission
parameters. Depending on the bandwidth of the system, as well as the subcarrier
spacing, a communication channel may appear sparse or dense in either the delay
or the Doppler domain. Hence, it is important to choose comparable parameters.
We now define hypothetical parameters for a \ac{OTFS} transmission system at
\ac{mmWave} frequencies. We define those parameters in
Tab. \ref{tab:system-parameters}. The system has a center frequency of $60\,$GHz.
For bandwidths, we consider $5$, $40$ and $120\,$MHz systems. On these, we use
$64$ subcarriers. Finally, we analyze different number of time aggregations in
order to investigate the limits OFTS performance.

\section{Measured V2V Channels}
\subsection{Measurement Campaign}
A detailed description of the measurement campaign and the measurement setup is
found in~\cite{zochmannmeasured}. For ease of understanding, the 
key parameters of the campaign is shown in Tab. \ref{tab:sounding-parameters}, and
outlined below.

The investigated scenario is close to overtaking scenarios passing a platoon.
Transmitter and receiver are placed next to an urban road, and the channel is
measured while cars pass by. The beginning of the measurement range is equipped
with a light-barrier that indicates when a new vehicle starts to pass by, which
automatically triggers a measurement.
The sample rate at the receiver is $600$\,Msamples/s. A multitone sequence is employed with $N=121$ carriers to approximately achieve a tone spacing of
$5$\,MHz. Due to the anti-aliasing filter, we avoid the cut-off region and only
transmit the sounding sequence at the $N_s=101$ center tones. Thereby, an
effective sounding bandwidth of $510$\,MHz is achieved. The output of our
channel sounder is the calibrated time-variant transfer function $\mtx{H}[t,f]$.

\begin{table}[t]
	\caption{Channel sounding measurement parameters}\label{tab:sounding-parameters}
	
	\centering
	\begin{tabular}{r|l}\toprule
		Parameter		&  Value \\ 
		\midrule
		Center frequency 	& $60$\,GHz \\ 
		Subcarrier spacing & $4.96$\,MHz  \\ 
		Number of subcarriers	& $102$ \\ 
		Snapshot rate	& $129.1$\,$\mu$s \\ 
		Delay resolution 	& $1.96$\,ns \\ 
		Recording time	& $720$\,ms \\ 
		\bottomrule
	\end{tabular} 
\end{table}
\subsection{Delay-Doppler Interpolation}
The recorded channel measurements store the results in a frequency-time grid
$\mtx H[f,t]$. However, to execute Eq. \ref{eq:rx}, we have to resample in
delay- and Doppler domain to match the simulation settings.
The total measurement bandwidth is $500\,$MHz, while the subcarrier spacing is
$5\,$MHz. We define these quantities as upper and lower bound of possible system
bandwidths.
To adapt the dimensions, we first ensure the new matrix $\mtx H'$ that has the same time dimension
$T$, but only uses a subset of $f'$ frequency rows of $\mtx H$. In this way, we
achieve the desired bandwidth. Then, we introduce the centered, unitary \ac{DFT}
matrix $\mtx F$. We calculate a delay-time representation $\mtx G$ via
\begin{equation}
  \mtx G = \begin{bmatrix}
    \mtx F_{f'\times f'}^H \mtx H' \\
    \mtx 0^{N-f' \times T}
  \end{bmatrix} .
\end{equation}  
By appending $N-f'$ rows of all zeros, we ensure that the system has the correct
number of subcarriers. Finally, we linearly interpolate between two consecutive
snapshots, to get the desired snapshot repitition rate. We consider the linear
interpolation to be of high quality, as the correlation coefficient between
consecutive snapshots, defined in our case as$\rho(\mtx a, \mtx b)
=\tfrac{\Re(<a,b>)}{\|a\|_2\|b\|_2},
$ is close to $0.95$. Thus, the samples are
highly correlated, and simple interpolation yields good performance.

\section{Performance Evaluation}\label{sec:results}
\begin{figure}
  \centering
  \subfloat[BW$ = 5\,$MHz\label{fig:perf64-a}]{ \includegraphics[width=\columnwidth] {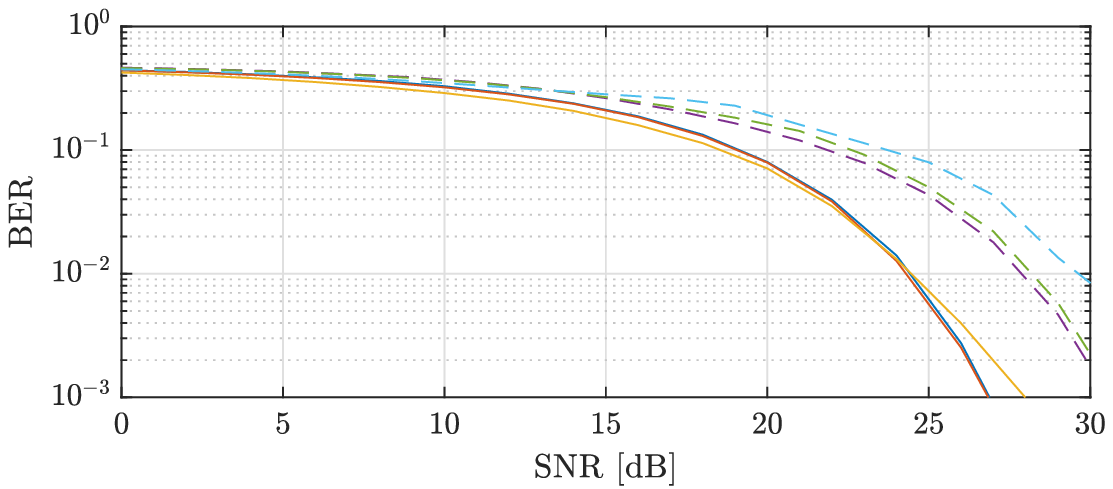}
  }\\\vspace{-1em}
  \subfloat[$BW = 40\,$MHz\label{fig:perf64-b}]{
    \label{fig:per64BW40}\includegraphics[width=\columnwidth]{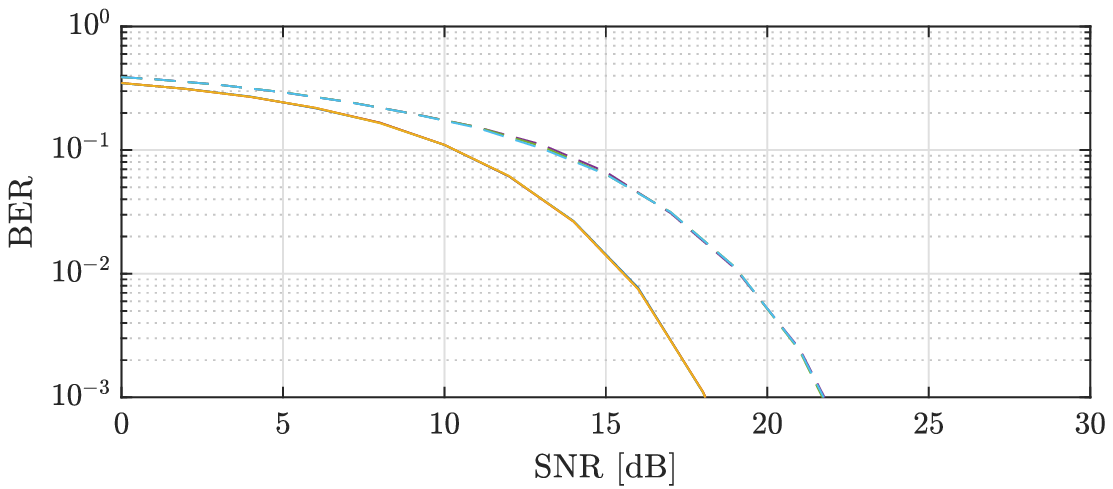}
  }
  \\\vspace{-1em}
  \subfloat[BW$ = 120\,$MHz\label{fig:perf64-c}]{
    \label{fig:per64BW120}\includegraphics[width=\columnwidth]{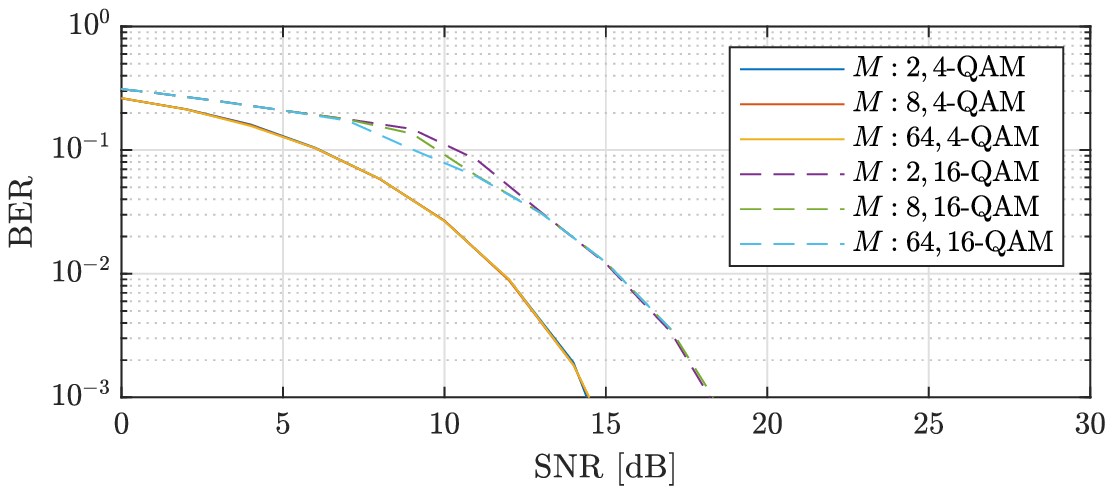}
  }\\\vspace{-1em}
  \subfloat[Synthetic channel\label{fig:perf64-d}]{
    \label{fig:per64BW40synCh}\includegraphics[width=\columnwidth]
    {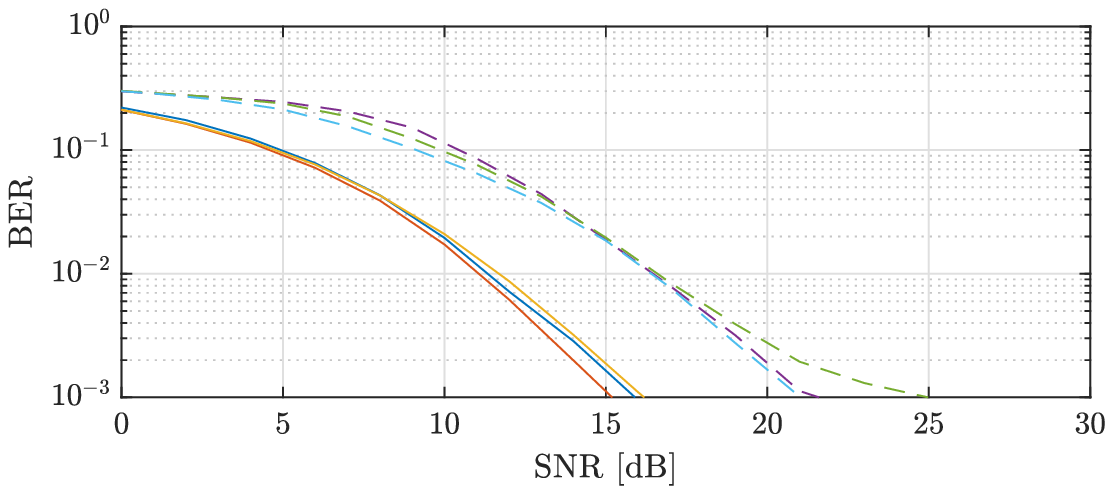}}
  \caption
  {Performance results for N=64 subcarriers. The legend is valid for all given subplots.}
  \label{fig:perf64}
\end{figure}

We conduct the bit error performance evaluation using the channel measurements
and the settings in Tab. \ref{tab:system-parameters}. 
The performance evaluation is done by comparing bit error rate (BER) at various
levels of \ac{SNR}. Here we use the SNR definition of expected energy per bit over noise power, $\text{SNR} = \tfrac{E_b}{N_0}$.

\begin{figure}
  \centering
  \includegraphics[width=\columnwidth]{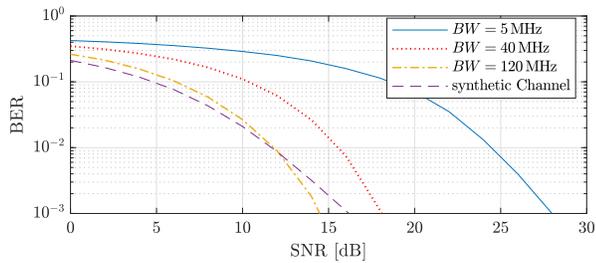}
  \caption{Comparison of different bandwidths, 64 Doppler taps, 4-QAM.}
  \label{fig:per64_40vs120}
\end{figure}

We simulate transmissions for
various OTFS configurations. As channel, we use a measurement trace where a
\ac{SUV} was passing by, while transmitter and receiver had line-of-sight connections.
For comparison, we also use
the synthetic channel described in \cite{viterbo-decoder}. The synthetic channel
has four Rayleigh-distributed taps in delay-Doppler domain, specifically the taps
have offset-indices of $\{(0,0), (1,1),(2,2),(3,3)\}$ in delay-Doppler domain, with equal
power across the taps. As they are defined in
terms of their indices and not absolute offsets, they are independent of the used bandwidth.
Figs. \ref{fig:perf64-a}, \ref{fig:perf64-b}, \ref{fig:perf64-c} show the achieved bit error rates
on the measured channel with different system bandwidths. Fig. \ref{fig:perf64-d} on
the other hand shows the performance over the synthetic channel.
Both bandwidths of $40$ and $120\,$MHz show performances that are independent of
the number of Doppler taps used in the \ac{OTFS} configuration. This can be
explained easily, as $\Delta\nu$, the lowest resolvable Doppler shift, given by
Eq. \ref{eq:condition} is
$9765.5$\,Hz in the case of $B=40$\,MHz. Meanwhile, a relative speed of
$50\,$km/h only translates to $2778\,$Hz Doppler shift. Thus, the channel is
completely compressed into one Doppler slot, and no diversity can be exploited without significant increment of N.
For the same parameters but at $5\,$MHz bandwidth, the Doppler resolution
becomes $1220.7$\,Hz. Thus, as can be seen, there is an observable gain in using
\ac{OTFS}. However, this comes with a severe penalty. The low bandwidth makes
the channel highly dense, and the overall achievable bit error rate performs
badly. In comparison, the synthetic channel demonstrates a visibly more well
behaved scenario.
Fig. \ref{fig:per64_40vs120} shows a direct performance comparison between the
different bandwidth constellations and the synthetic channel. The comparison is
done with $M=64$, and modulation scheme 4-QAM. The comparison demonstrates
that the synthetic channel is a strongly optimistic estimation of the severeness
of actual channels.
One possible mitigation for this is to use large bandwidths and drastically
increase $N$. However, due to the complexity of the iterative algorithm, this
results in computationally prohibitively slow decoding steps.

\section{Conclusions}
We provide performance simulations for \ac{OTFS} based on measured vehicular
channels. Our results show that using OTFS can provide performance gains by exploiting two-dimensional modulation concepts. However, the used system bandwidth and Doppler resolution are linked via the modulation parameters $M$ and $N$. These links remove degrees of freedom, which can stop the system from exploiting diversity in one of the considered domains.
For channel estimation to benefit from OTFS, the channel has to be sparse to keep complexity low, yet spread out in both delay and Doppler
domains. However, design choices that spread the channel in both domains run the risk
of either increasing the denseness of the channel, or increasing the symbol dimension to
computationally prohibitive sizes. On the other hand, sparse channels may become
one-dimensional, removing the diversity gains.
One solution to this problem may be to go for computationally more efficient receiver structures that allow denser subcarrier spacings, as well as applying, channel coding. In any case, measures have to be taken to ensure performance gains.
\section*{Acknowledgements}
{\footnotesize
This work has been conducted within the DARVIS project that has received funding from the Austrian Aeronautics Research and Technology Program TAKEOFF under grant agreement No 867400.
}
\bibliographystyle{IEEEtran} 
\input{./latex-base-files/acronyms}
\bibliography{./latex-base-files/library,./latex-base-files/temp}

\end{document}

%% file: latex-base-files/packages.tex
  \usepackage{tikz}
  \usepackage{pgfplots}
  \usepackage[caption=false,font=footnotesize]{subfig}
  \usepackage{multirow}
  \pgfplotsset{compat=1.12}
  \usepackage[T1]{fontenc} 
  \usepackage[cmintegrals]{newtxmath}
  \usepackage[nolist]{acronym}
  \usepackage[capitalize,noabbrev]{cleveref}


%% file: latex-base-files/acronyms.tex
\begin{acronym}
    \acro{ADC}{Analog-Digital Converter}
    \acro{AIC}{Akaike Information Criterion}
    \acro{AGC}{Automatic Gain Control}
    \acro{AWGN}{Additive White Gaussian Noise}
    \acro{C-ITS}{Cooperative Intelligent Transport Systems}
    \acro{c-LASSO}{Complex \ac{LASSO}}
    \acro{CAM}{Cooperative Awareness Message}
    \acro{CCA}{Clear Channel Assessment}
    \acro{CDF}{Cumulative Distribution Function}
    \acro{CP}{Cyclic Prefix}
    \acro{CRC}[\textrm{CRC}]{Cyclic Redundancy Check}
    \acro{CRC}{Cyclic Redundancy Check}
    \acro{CSI}{Channel State Information}
    \acro{CSMA/CA}{Carrier Sense Multiple Access with Collision Avoidance}
    \acro{CVIS}{Cooperative Vehicle-Infrastructure Systems}
    \acro{DFT}{Discrete Fourier Transform}
    \acro{DSFT}{Discrete Symplectic Fourier Transform}
    \acro{ECDF}{Empirical Cumulative Distribution Function}
    \acro{ETSI ITS-G5}{\acl{ETSI} \ac{ITS} at $5.9\,$GHz}
    \acro{ETSI}{European Telecommunications Standards Institute}
    \acro{FCS}{Frame Check Sequence}
    \acro{FPGA}{Field-Programmable Gate Array}
    \acro{GPS}{Global Positioning System}
    \acro{GSCM}{Geometry-based Stochastic Channel Model}
    \acro{HMM}{Hidden Markov Model}
    \acro{IDFT}{Inverse \ac{DFT}}
    \acro{IDM}{Intelligent Driver Model}
    \acro{ISI}{Intersymbol Interference}
    \acro{ITS}{Intelligent Transport Systems}
    \acro{LASSO}{Least Absolute Shrinkage and Selection Operator}
    \acro{LOS}{Line-of-Sight}
    \acro{LTE}{Long Term Evolution}
    \acro{MAC}[\textrm{MAC}]{Medium Access Control}
    \acro{MAC}{Medium Access Control}
    \acro{MCS}{Modulation and Coding Scheme}
    \acro{MIMO}{Multiple-Input Multiple-Output}
    \acro{MMSE}{Minimum Mean Squared Error}
    \acro{mmWave}{Millimeter Wave}
    \acro{MP}{Message Passing}
    \acro{MPC}{Multipath Component}
    \acro{MSE}{Mean Squared Error}
    \acro{NI}{National Instruments}
    \acro{NLOS}{Non Line-of-Sight}
    \acro{NLOS}{Non-Line of Sight}
    \acro{OAM}{Orbital Angular Momentum}
    \acro{OEW}{Open-Ended Waveguide}
    \acro{OFDM}{Orthogonal Frequency Division Multiplexing}
    \acro{OTFS}{Orthogonal Time Frequency Space}
    \acro{PCMA*}{Predictive Congestion Minimization in Combination with an A* based router}
    \acro{pdf}{probability density function}
    \acro{PDP}{Power Delay Profile}
    \acro{PDR}{Packet Delivery Rate}
    \acro{PER}{Packet Error Rate}
    \acro{PHY}[\textrm{PHY}]{Physical Layer}
    \acro{pmf}{probability mass function}
    \acro{PSD}{Power Spectral Density}
    \acro{QPSK}{Quaternary Phase Shift Keying}
    \acro{RMS}{Root Mean Squared}
    \acro{RSSI}{Received Signal Strength Indicator}
    \acro{RSSI}{Received Signal Strength Indicator}
    \acro{RSU}{Roadside Unit}
    \acro{SDR}{Software Defined Radio}
    \acro{SIR}{Signal-to-Interference Ratio}
    \acro{SNR}{Signal-to-Noise Ratio}
    \acro{SUV}{Sports Utility Vehicle}
    \acro{USRP RIO}{\aclu{USRP} with Reconfigurable I/O}
    \acro{USRP}{Universal Software Radio Peripheral}
    \acro{V2I}{Vehicle-to-Infrastructure}
    \acro{V2V}{Vehicle-to-Vehicle}
    \acro{V2X}{Vehicle-to-Everything}
    \acro{VANET}{Vehicular Ad-Hoc Network}
    \acro{WSSUS}{\aclu{WSS} Uncorrelated Scattering}
    \acro{WSS}{Wide-Sense Stationary}
    \acro{ZF}{Zero Forcing}
\end{acronym}

%% file: conference_041818.bbl
\begin{thebibliography}{10}
\providecommand{\url}[1]{#1}
\csname url@samestyle\endcsname
\providecommand{\newblock}{\relax}
\providecommand{\bibinfo}[2]{#2}
\providecommand{\BIBentrySTDinterwordspacing}{\spaceskip=0pt\relax}
\providecommand{\BIBentryALTinterwordstretchfactor}{4}
\providecommand{\BIBentryALTinterwordspacing}{\spaceskip=\fontdimen2\font plus
\BIBentryALTinterwordstretchfactor\fontdimen3\font minus
  \fontdimen4\font\relax}
\providecommand{\BIBforeignlanguage}[2]{{%
\expandafter\ifx\csname l@#1\endcsname\relax
\typeout{** WARNING: IEEEtran.bst: No hyphenation pattern has been}%
\typeout{** loaded for the language `#1'. Using the pattern for}%
\typeout{** the default language instead.}%
\else
\language=\csname l@#1\endcsname
\fi
#2}}
\providecommand{\BIBdecl}{\relax}
\BIBdecl

\bibitem{Blaz1809:Model}
T.~Blazek, E.~{Z{\"o}chmann}, and C.~F. {Mecklenbr{\"a}uker}, ``Model order
  selection for {LASSO} fitted millimeter wave vehicular channel data,'' in
  \emph{Proc. of 29th Annual International Symposium on Personal, Indoor, and
  Mobile Radio Communications (PIMRC)}.\hskip 1em plus 0.5em minus 0.4em\relax
  Bologna, Italy: IEEE, Sep 2018, pp. 1--5.

\bibitem{Mecklenbrauker2011}
C.~F. Mecklenbr\"auker, A.~F. Molisch, J.~Karedal, F.~Tufvesson, A.~Paier,
  L.~Bernado, T.~Zemen, O.~Klemp, and N.~Czink, ``{Vehicular Channel
  Characterization and Its Implications for Wireless System Design and
  Performance},'' \emph{Proc. of the IEEE}, vol.~99, no.~7, pp. 1189--1212, Jul
  2011.

\bibitem{naik2019ieee}
G.~Naik, B.~Choudhury, and J.-M. Park, ``Ieee 802.11 bd \& 5g nr v2x: Evolution
  of radio access technologies for v2x communications,'' \emph{IEEE Access},
  vol.~7, pp. 70\,169--70\,184, 2019.

\bibitem{monk2016otfs}
A.~Monk, R.~Hadani, M.~Tsatsanis, and S.~Rakib, ``Otfs-orthogonal time
  frequency space,'' \emph{arXiv preprint arXiv:1608.02993}, 2016.

\bibitem{hadani2017orthogonal}
R.~Hadani, S.~Rakib, A.~Molisch, C.~Ibars, A.~Monk, M.~Tsatsanis, J.~Delfeld,
  A.~Goldsmith, and R.~Calderbank, ``Orthogonal time frequency space (otfs)
  modulation for millimeter-wave communications systems,'' in \emph{2017 IEEE
  MTT-S International Microwave Symposium (IMS)}.\hskip 1em plus 0.5em minus
  0.4em\relax IEEE, 2017, pp. 681--683.

\bibitem{viterbo-decoder}
P.~{Raviteja}, K.~T. {Phan}, Y.~{Hong}, and E.~{Viterbo}, ``Interference
  cancellation and iterative detection for orthogonal time frequency space
  modulation,'' \emph{IEEE Transactions on Wireless Communications}, vol.~17,
  no.~10, pp. 6501--6515, Oct 2018.

\bibitem{zemen2018iterative}
T.~Zemen, M.~Hofer, D.~Loeschenbrand, and C.~Pacher, ``Iterative detection for
  orthogonal precoding in doubly selective channels,'' in \emph{2018 IEEE 29th
  Annual International Symposium on Personal, Indoor and Mobile Radio
  Communications (PIMRC)}.\hskip 1em plus 0.5em minus 0.4em\relax IEEE, 2018,
  pp. 1--7.

\bibitem{TUW-279699}
T.~Blazek, H.~Groll, S.~Pratschner, and E.~Z{\"o}chmann, ``Vehicular channel
  characterization in orthogonal time-frequency space,'' in \emph{Proceedings
  of the IEEE ICC 2019}, IEEE, Ed., 2019, talk: IEEE International Conference
  on Communications (ICC 2019), Shanghai, China; 2019-05-20 -- 2019-05-24.

\bibitem{murali2018otfs}
K.~Murali and A.~Chockalingam, ``On otfs modulation for high-doppler fading
  channels,'' in \emph{2018 Information Theory and Applications Workshop
  (ITA)}.\hskip 1em plus 0.5em minus 0.4em\relax IEEE, 2018, pp. 1--10.

\bibitem{ramachandran2018mimo}
M.~K. Ramachandran and A.~Chockalingam, ``Mimo-otfs in high-doppler fading
  channels: Signal detection and channel estimation,'' in \emph{2018 IEEE
  Global Communications Conference (GLOBECOM)}.\hskip 1em plus 0.5em minus
  0.4em\relax IEEE, 2018, pp. 206--212.

\bibitem{raviteja2019embedded}
P.~Raviteja, K.~T. Phan, and Y.~Hong, ``Embedded pilot-aided channel estimation
  for otfs in delay--doppler channels,'' \emph{IEEE Transactions on Vehicular
  Technology}, vol.~68, no.~5, pp. 4906--4917, 2019.

\bibitem{8746382}
G.~D. {Surabhi}, M.~K. {Ramachandran}, and A.~{Chockalingam}, ``Otfs modulation
  with phase noise in mmwave communications,'' in \emph{2019 IEEE 89th
  Vehicular Technology Conference (VTC2019-Spring)}, April 2019, pp. 1--5.

\bibitem{khammammetti2018otfs}
V.~Khammammetti and S.~K. Mohammed, ``Otfs-based multiple-access in high
  doppler and delay spread wireless channels,'' \emph{IEEE Wireless
  Communications Letters}, vol.~8, no.~2, pp. 528--531, 2018.

\bibitem{zochmannmeasured}
E.~Z{\"o}chmann \emph{et~al.}, ``Measured delay and {Doppler} profiles of
  overtaking vehicles at 60 {GHz},'' in \emph{Proc. of the 12th European
  Conference on Antennas and Propagation (EuCAP)}, London, Great Britain, 2018,
  pp. 1--5.

\bibitem{blazek2018millimeter}
T.~Blazek, E.~Z{\"o}chmann, and C.~Mecklenbr{\"a}uker, ``Millimeter wave
  vehicular channel emulation: A framework for balancing complexity and
  accuracy,'' \emph{Sensors}, vol.~18, no.~11, p. 3997, 2018.

\end{thebibliography}
